\documentclass[12pt]{article}
\usepackage{graphicx,times,achemso,bm,amssymb}

\title{Aging and metastability of monoglycerides in hydrophobic solutions}
\author{C.H.~Chen, 
        and E.M.~Terentjev   
\\ {  } \\
{ \small Cavendish Laboratory, University of Cambridge}
\\ {\small J J Thomson Avenue, Cambridge CB3 0HE, U.K.}
 }
\begin{document}
\maketitle
\renewcommand{\thefootnote}{\fnsymbol{footnote}}
\noindent The aging of aggregated structures of monoglycerides in hydrophobic medium is described by a set of different techniques. Polarized microscopy was used to study the mesomorphic behavior as a function of time. Differential scanning calorimetry was utilized to quantitatively monitor changes in the latent heat in different phase transformations that take place in the aging system. Infrared spectroscopy was applied to detect the formation of hydrogen bonding between surfactants. The X-ray diffraction patterns fingerprinted the molecular arrangement in different emerging phases. Infra-red spectroscopy was used to monitor the state of hydrogen bonding in the system.
We conclude that in both inverted-lamellar and sub-alpha crystalline phases, monoglyceride molecules inevitably lose their emulsified ability in the hydrophobic solutions through the gradual change in hydrogen bonding patterns. On aging, the formation of intermolecular hydrogen bonding between glycerol groups causes the segregation of chiral ($D$ and $L$) isomers within the bilayers. Therefore all  structures were eventually forced to reorder into the beta-crystalline state, distinguishing between the $D$ and $L$ layers. Accordingly, the highly ordered packing of aged structures weakened the emulsifying ability and finally leaded the collapse of the percolating gel network.

\newpage
\section{Introduction}

Monoglycerides (MG) are lipid molecules consisting of a single fatty acid esterified with a 1-hydroxy glycerol group. They are distinguished by the length of carbon chain. In this paper we focus on a particularly common surfactant C18. Unlike typical non-ionic surfactants, MG is an optically active molecule that exists in two chiral isomers: $D$ and $L$.\cite{Agt98, Mohwald96} Concentrated mixtures of MG in solvents form the cream-like materials widely used both in personal products and food industries.\cite{Veeman05, Hendricx98} Attractive features of MG dispersed oil include two main factors: uniquely, these molecules form an elastic gel network on initial aggregation from the isotropic into the lamellar phase, which then retains its mechanical stability on further cooling.\cite{first} In practical applications, partly due to this mechanical stability and partly to their biocompatibility, MG/oil mixtures could be used as healthy substitutes for butter. Due to the absence of water component, MG/oil systems are not easily susceptible to microbial infection and autoxidation, and thus can be stored at room temperature without the need for preservative additives.\cite{Shimoni04a ,Shimoni04b, Shimoni07, Sagalowicz07} Beyond their use in various applications, in general, aggregated surfactant solutions form a particular class of materials that develop a network of morphological features to retain the solvent.\cite{Sein02, Florence99} For these reasons, the detailed knowledge of MG/oil systems is important both for scientific advancement and technological implementation.

MG systems have been studied systematically for many years. The structure of pure MG was firstly described by Larsson in 1966 \cite{Larsson66} and was later reviewed by Small (1986)\cite{book_Small}, Larsson (1994)\cite{Larsson94}, and Krog (2001)\cite{book_Garti}. The polymorphic behavior of MG/water is well established. On temperature decrease, the emerging could be divided into three groups.\cite{Mohwald96 ,Larsson68} In the high-temperature region, the materials is in the isotropic fluid phase (which may involve some micellar aggregation). At a lower temperature, the lamellar phase exists over a wide range of concentrations.\cite{Agt98} According to the literature, below the Krafft temperature, $T_{K}$, the lamellar phase transforms into the alpha-crystalline phase which contained the hexagonal surfactant packing, characterized by a single X-ray short spacing at 4.18\AA.\cite{book_Aljandro, Marangoni07c}  Results of our recent work \cite{first} indicate that the same structural feature is seen in the inverse-lamellar phase of MG in the oil matrix, where the characteristic 4.17\AA spacing arises from the hexagonal ordering of glycerol heads in the tightly compressed mid-bilayer planes (while the aliphatic tails remained disordered, although laterally densely packed). Since the characteristically concentration-independent Krafft point $T_{K}$ involves crystallization on demixing, we are concerned about the structural interpretation of the alpha-phase in aqueous systems -- however, this is not the subject of the present article, which deals with purely non-aqueous systems.
A detailed study of the head-group hydration of revealed that there were three water molecules bonded to one surfactant molecule in the lamellar phase.\cite{Tiddy93} In the alpha-crystalline state, these three bonds became one and, as a result, weakened the emulsification ability. Moreover, the stability of the MG gel network was mainly governed by the van der Waals attraction force and short-range repulsive force between bilayers.\cite{Claesson90, Simon89} The repulsive pressure between bilayers in the alpha crystalline phase was significantly smaller than the pressure in the lamellar phase.

The initial alpha-crystalline gel is metastable and readily transforms into an anhydrous MG crystal,
identified as beta-crystalline state (often called the ``coagel''),\cite{Agt98, Veeman05} which has a higher melting point and is characterized by several wide-angle X-ray reflections characterizing
  short distances of the unit cell, with the strongest line corresponding to 4.5-4.6\AA.\cite{book_Garti, book_Small} A coagel state of MG in water is due to hydrogen bonds establishing within head groups in bilayers, which in turn lead to a further crystallization of aliphatic tails.\cite{Agt98} On a long time scale of aging, the D- and L- isomers of chiral MG gradually separate within crystalline bilayers, leading to more dense packing and expulsion of water. Sedimentation of solid in this phase then takes places.

\begin{figure}
\begin{center}
\resizebox{0.5\textwidth}{!}{\includegraphics{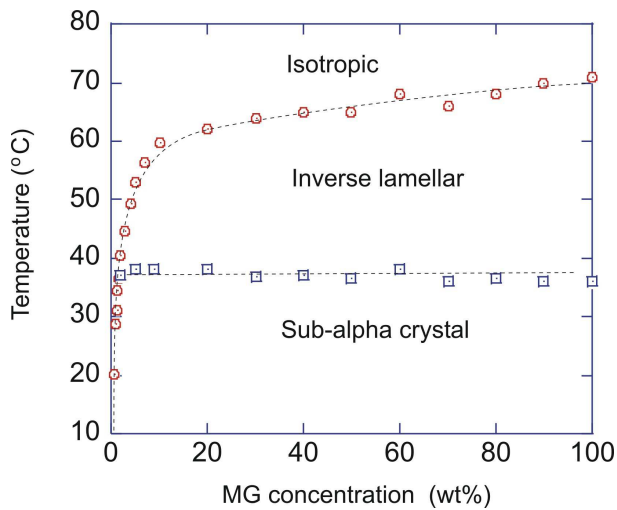}} \caption{ The phase diagram of MG in oil showed three phases between 0$^{\circ}$C and 80$^{\circ}$C: isotropic, inverse lamellar, and sub-alpha crystalline phases. The crystallization temperature, $T_K$ is characteristically independent of concentration. } \label{fig1}
\end{center}
\end{figure}

Compared with the aqueous systems, MG in a hydrophobic solvent is much less studied. The rheology and storage properties have been reported in a series of papers by Shimoni et al\cite{Shimoni04a, Shimoni04b, Shimoni07}. However due to the absence the confident structure descriptions, the mechanism of aging phenomena was still unclear. There are many examples of the use of ternary MG/water/oil systems.\cite{otto06,Marangoni07c,Marangoni06b,Marangoni07a,Fernandez01,Ian03,Larsson90} However the presence of water would dominate the phase behavior of aggregating MG even at the lowest water content (0.5\% w/w).\cite{first} In our recent work, we have studied the MG C18 in oil in the absence of water. Having checked the drying oil condition, and the mixtures of C18/C16 at different concentrations, we have established a reasonably universal phase diagram, Fig.\ref{fig1}. Due to the difference of hydrogen bonding patterns, MG/oil showed the  phase behavior very different from the aqueous systems. Below the isotropic-lamellar transition temperature, $T_{L}$, the inverse lamellar phase was formed, as expected in the hydrophobic system. However, due to the unique size ratio between the glycerol head and the lateral area of aliphatic chain in the fully extended dense brush, the inverse lamellar bilayers have a very definite hexagonal ordering. We have concluded that this is a 2-dimensional dense packing of glycerol heads, compressed in the middle of each bilayer. One might suggest that the hexagonal order arises from laterally densely packed aliphatic tails, e.g. \cite{Marangoni07c} However if we make a comparison with other densely grafted polymer brushes -- and more importantly, other single-aliphatic tail surfactants, none of which have such a hexagonal packing, the much more likely scenario is that the heads are densely packed in 2D planes and then the tails (in the fully extended due to the brush density, but nevertheless molten state) simply follow that effective grafting pattern.
 The wide-angle Xray reflection at 4.17\AA corresponds to the closest distance of approach of glycerol groups in a plane. We have concluded that the second, ``twin'' peak at 4.11\AA, is the characteristic distance between the neighboring heads in two planes of the bilayer. Continuously cooling down below the crystallization point of the surfactant, the lower-temperature phase transition emerges. Characteristically, this crystallization temperature does not depend on MG concentration, which suggests that this is the Krafft demixing point, $T_K$. The lateral hexagonal packing transformed into a structure analogous to the ``sub-alpha'' crystalline form seen in aqueous systems, with orthorhombic chain packing in the unit cell, characterized by strong X-ray reflection at 4.17\AA and several reflections between 4.06 to 3.6\AA. \cite{first} The reason why the ordinary (in aqueous systems) alpha-crystal state is not observed in the hydrophobic environment is that the highly 2D-ordered glycerol heads promote a higher degree of order on the subsequent crystallization of aliphatic tails,
 thus bypassing the alpha-crystalline state which still has hydrated glycerol groups on the outside of each bilayer.

This unique structural feature of the inverse lamellar phase is the origin of the well-known rheological feature of MG/oil systems: they form an elastic gel at a high temperature, immediately below the isotropic-lamellar transition. The structured bilayers have much less flexibility and thus form a percolating scaffold that can resist the macroscopic stress applied to the system. On entering lower-temperature crystalline phase the elastic modulus of the gel was shown not to change significantly, although the yield stress of course does change due to the different bending rigidity of bilayers.

In the present work we investigate the aging processes of MG/oil in both inverse-lamellar and sub-alpha crystalline phases. The textures and their evolution were observed through polarized microscopy. During aging we find that the network structure breaks down into the isolated beta-crystalline clusters dispersed in a continuous oil phase. Therefore the lamellar network scaffold breaks down and phase separation between solid MG crystals and liquid oil takes place; as a result the system loses its rheological characteristics of a gel. This aging process was quantitatively analyzed by a comparison of melting latent heat of transitions in aged and fresh materials, using differential calorimetry (DSC). The structure of beta-crystal phase was characterized by X-ray diffraction. In order to understand the mechanism of aging, the time evolution of hydrogen bonding states was monitored by infrared spectroscopy. By comparing the relevant literature data and our infrared absorption peaks, we make conclusions about the formation of intermolecular hydrogen bonds during the aging process. Putting together results of several different experiments, we conclude that the MG/oil mixtures in both inverse-lamellar and sub-alpha crystalline phases are in fact metastable. Such a conclusion is expected for the sub-alpha crystalline phase (and corresponds to what takes places in aqueous systems), however, the metastability of the lamellar phase is an unusual finding. In both cases, with the passage of time the MG molecules rearrange into a true stable structure of the beta-crystal, with alternating $D$ and $L$ MG layers. Accordingly, the highly ordered packing of aged structures weakened the emulsifying ability and causes the collapse of the gel-network.

\section{Experimental details \label{sec:1}}


Distilled saturated MG were purchased from the Palsgaard A/S (Denmark). The sample contained 97\% 1-monoglyceride and the fatty acid chain length composition was 93\% C18 and 7\% C16. The remainder consisted  of 1.1\% diglycerides and less than 1\% triglycerides. The hydrophobic solvent used in the bulk of this work was hazelnut oil from Provence, France, which variety contains approximately 80\% oleic and 20\% linoleic acids with low quantities of MG.\cite{oil}. This oil crystallizes at a temperature below $-23^{\circ}$C. Before the testing the hazelnut oil was heated to
$120^{\circ}$ for several hours to keep its drying condition. The choice of this particular oil is partially dictated by practical application. Later on, in the discussion of our results, one may question the extent to which the small impurity of the MG C18, and of the hazelnut oil, was important. From the earlier studies of equilibrium ordering \cite{first} we are clear that this is not important at all and the conclusions are quite universal. As in that study, here we have also carried out parallel tests on MG in n-tetradecane (the ``model oil'') and found no significant changes in either infrared studies of hydrogen bonding, or in calorimetric characterization of phase transitions (apart from a small shift of the lamellar transition temperature).

For this study, MG was mixed with the oil at a fixed concentration of 10\% w/w. To ensure proper mixing, the solutions were placed on a magnetic stirrer heating plate at a constant temperature of 100$^{\circ}$C. In order to follow the aging in both sub-alpha crystalline and inverse lamellar phases, the samples were stored at 26$^{\circ}$C and 50$^{\circ}$C, respectively.


A Zeiss Axioplan microscope with crossed polarizers was attached to a Linkam TP91 hot stage unit. Samples of the 10\% w/w MG/hazelnut oil were placed between a glass slide and a cover slip, with the thickness of the sample less than 0.5mm. The sample was heated to 100${}^{\circ}$C and annealed at this temperature for five minutes to eliminate thermal history. To observe the time evolution, samples were cooled from the isotropic phase and kept below relevant the transition temperature for two weeks.


Heat exchanges involved in a phase transition yielded exothermic or endothermic peaks that were recorded in a differential scanning calorimeter (DSC). From these measurements the transition temperatures and the latent heat could be accurately estimated. A Perkin-Elmer power-compensated Pyris 1 differential scanning calorimeter equipped with an intracooler 2P was used. The samples for different periods of time were placed in the pans and heated to 100$^{\circ}$C from 26$^{\circ}$C (sub-alpha crystalline phase), or from 45$^{\circ}$C (inverse lamellar phase) to record the temperature and the latent heat on melting. After that, samples were held for 1 min at 100$^{\circ}$C and cooled to 0$^{\circ}$C, to assess the stability of the aged phase to melting nucleation. Each sample was measured in 40ul sealed aluminum pans at a scanning rate of 20$^{\circ}$C/min. The choice of this relatively high rate of temperature change was guided by the fact that we saw no qualitative differences in the transitions on testing at 10, 5 or 1$^{\circ}$C/min -- but the sensitivity of DSC technique decreases dramatically and our errors in determining the coagel index increased. All aging effects described here take a much longer time, while the initial lamellar ordering of the surfactant occurs sufficiently fast for us not to see a significant difference at varied cooling rates.


To determine the characteristics of hydrogen bonds, infrared absorption spectra were recorded
employing a GALAX series 4020 FT-IR spectrometer (Thermal
Electron Corporation, Waltham, MA, USA). The spectrometer was continuously purged excluding $CO_{2}$ from the sample holder box. The samples were placed in an IR cell between two $NaCl$ windows. The temperature was regulated by placing the cell in a thermal stated holder controlled by a temperature controller (Eurotherm, SPECAC). To observe the thermal effect, the samples were prepared and the evolution of spectra was recorded from 20 to 100 $^{\circ}$C with the interval of 10$^{\circ}$C, and then on cooling from 100$^{\circ}$C to 26$^{\circ}$C. After the heating treatment the experiment was
recorded continuously at room temperature every 24 hours for 14 days to monitor the aging process.


X-ray scattering patterns were recorded using a copper rotating anode generator (Rigaku-MSC Ltd) equipped with X-ray optics by Osmic Ltd. Before recording the X-ray diffraction, the samples were stored at 26 $^{\circ}$C (crystalline phase), or at 50 $^{\circ}$C (lamellar phase) for 14 days. The distance between the detector and the sample was set 300mm, giving a maximum resolution of 3.36 ${\AA}$ at the edge of the diffraction pattern. Samples were held between a mica sheet of 0.1mm (supplied by Goodfellow, Cambridge, UK) and an aluminium plate. A metal substrate plate was used to ensure accurate heat transfer to the sample. The temperature was controlled by a home-made chamber and verified by a thermocouple.\cite{Goto88}

\section{The aging process}

\begin{figure}
\begin{center}
\resizebox{0.65\textwidth}{!}{\includegraphics{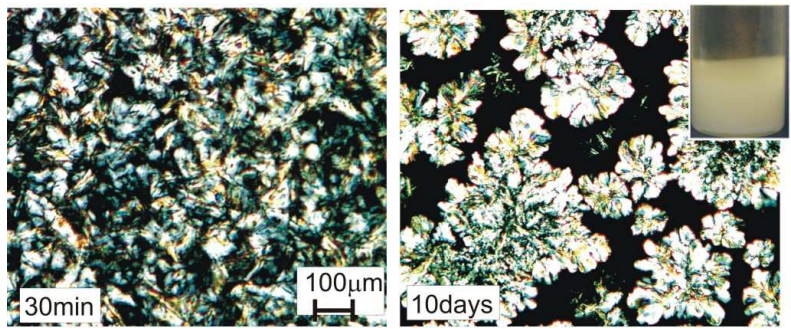}} \caption {At 26${}^{\circ}$C, MG initially aggregated into a lipid crystal network (elastic gel) that trapped the residual oil inside. After a 10-day evolution, the lamellar network was broken and crystalline regions aggregated together. The phases of oil and crystalline MG separate, which showed as a clear sedimentation boundary in the vessel (inset shows the 10-day old coagel).} \label{fig2}
\end{center}
\end{figure}

The aging of the textures in sub-alpha crystalline phase was observed by polarized optical microscopy. Fig.\ref{fig2} showed the typical microstructure at fixed temperature 26$^{\circ}$C, separated by a long period of aging. In order to record clear images, the sample thickness was less than 0.5mm. As always, the samples were annealed at 80${}^{\circ}$C and then cooled down to a certain temperature, after which the texture was record every 3 hours for two days and subsequently every 12 hours for two weeks. When the mixture is cooled to 26${}^{\circ}$C, MG initially form what we continue to call the ``sub-alpha crystal'' phase (by analogy with a similar structure in aqueous systems), which consists of dispersed crystalline bilayer plates making a polydomain percolating network throughout the sample. This texture shows bright birefringence under the crossed polarizers. The size of every elongated plate domain was roughly $100\mu$m in length and $25\mu$m in width. After a 10-day evolution (essentially slow re-crystallization and phase separation from the residual oil) the structure changes into solid crystal clusters dispersed in a continuous oil phase. In this case the gel-forming network is broken and the oil completely phase separated from the MG crystals, which then gradually sediment by gravity.

The aging process could be followed thermodynamically by introducing a dimensionless number called the ``coagel index'' (CI) which is the ratio of total latent heat of the transition in fresh and aged materials: CI$= \Delta H(\textrm{aged})/ \Delta H(\textrm{fresh})$. The aged structure of aggregated MG contains beta-crystals with a highly regular arrangement achieved through segregated molecular chirality. This caused the latent heat of the aged material to be higher than that of the freshly ordered system. Therefore the CI could be used to monitor the fraction of the aged
part in the sample by comparing the heat flow used to melt the freshly-ordered and the aged material.

\begin{figure}
\begin{center}
\resizebox{0.8\textwidth}{!}{\includegraphics{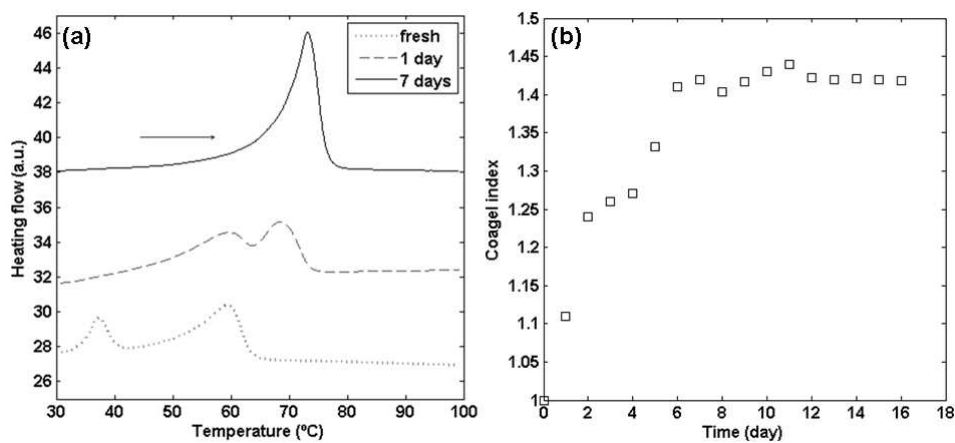}} \caption {(a) First heating of the fresh sample shows two transition peaks corresponding to the melting of sub-alpha crystal and the lamellar phase. After one day at 26$^{\circ}$C the DSC showed two peaks starting to coalesce. After five days the sample changed to form beta-crystals and exhibits only one melting peak at higher temperature. (b) CI increased gradually from 1.0 to 1.4 over approximately five days. } \label{fig3}
\end{center}
\end{figure}

The result of our DSC study shows that CI increases gradually with time the material spends at low temperature, and eventually reaches a saturation state. In order to observe the time evolution of CI, the samples in these experiments were stored at a fixed temperature (26$^{\circ}$C) after quenching into the initial sub-alpha crystalline state. The typical DSC scans are presented in Fig.\ref{fig3}(a). As found in the earlier studies,\cite{first} in the fresh sample one finds a sequences of two transition peaks between 0 and 100$^{\circ}$C . The first peak corresponds to the melting of the crystallized carbon chains of sub-alpha crystals and the second peak corresponded to the melting of inverted lamellar phase. After five days of aging at room temperature (26$^{\circ}$C), the two phase transitions displayed in the fresh sample coalesced into one.

The melting enthalpy of aged beta-crystalline samples was about 1.4 times the value obtained from melting the fresh samples (sub-alpha crystal). This 40\% difference, and the evolution of CI on aging, is shown in Fig.\ref{fig3}(b). The initial value of $CI=1$ indicates the sample melting from the sub-alpha state, while the saturation value of 1.4 suggests the better-ordered beta-crystalline state with a higher melting entropy. The time-evolution of CI shows that the sub-alpha phase is metastable and transformed into a more stable beta-crystal arrangement over approximately five days.

\begin{figure}
\begin{center}
\resizebox{0.75\textwidth}{!}{\includegraphics{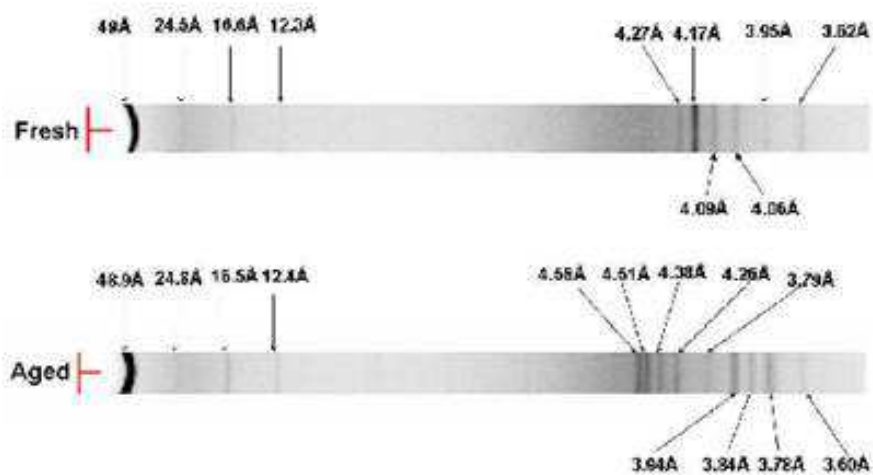}} \caption{Sub-alpha crystalline phase is metastable and transforms into the stable beta crystal form during storage at ambient temperature. In the beta crystal the alkyl chains are packed triclinically. The main spacing of glycerol head increases from 4.17 $\AA$ (sub-alpha crystal) to 4.55 $\AA$ (beta crystal). This is an evidence of the formation of intermolecular hydrogen bonds. }
\label{fig4}
\end{center}
\end{figure}

In order to describe the molecular arrangement during aging at room temperature (26$^{\circ}$C), X-ray diffraction was employed to distinguish between the structures of fresh and aged samples, see Fig.\ref{fig4}. The series of small angle reflections showed that both fresh and aged materials contained lamellar bilayers of approximately 49$extrm{\AA}$ thickness: the ratios of high-order small-angle reflections followed the characteristic sequence of $1, 1/2, 1/3, 1/4$. \cite{first} The wide angle diffraction reveals the differences. The fresh crystalline phase depicts sub-alpha crystalline ordering which is characterized by the orthorhombic unit-cell of chain packing with pronounced sequence of wide angle diffraction peaks at 4.27\AA-4.17\AA-4.09\AA-4.06\AA-3.95\AA-3.62\AA. The particularly strong reflection at 4.17\AA is the same as in the inverse-lamellar phase, which to us suggests that it originates from the 2D hexagonal packing of glycerol heads. After five days of aging the material transforms to a beta-crystal form. In this case the all-trans zigzags of the alkyl chains are parallel with each other and the chains pack triclinically. The beta-crystal form is characterized by two series of X-ray lines in the short-spacing region, 4.55\AA-4.51\AA-4.38\AA-4.26\AA-4.09\AA and 3.94\AA-3.84\AA-3.78\AA-3.60\AA.  One of these series (at wider angles) corresponds to the crystallized alkyl chains. We believe that the other series (at slightly smaller angles) reflects the correspondingly skewed periodic spacing between the glycerol heads in their planes. Note, that Fig.\ref{fig4} suggests that the glycerol head spacing was increased from 4.17$\textrm{\AA}$ to $4.55\textrm{\AA}$. This again is an indication of the formation of intermolecular hydrogen bonding.

During aging, intermolecular hydrogen bonds form between 1-hydroxy glycerol groups and force the molecules to rearrange their packing through separation of chiral isomers. Establishment of regular hydrogen bonds expanded the local volume of each molecular head in the layer, and
so increased the spacing between glycerol moieties, now mediated by the shared proton. This effect is analogous to what happens in the formation of ice from liquid water. Below the crystallization temperature, regular intermolecular hydrogen bonds form between water molecules. These intermolecular bonds drive the rearrangement of water molecules into the lattice, and also expand the local volume (so the spacing between water molecules increases and the material density falls). The details of hydrogen bonding between MG molecules will be discussed in the following section.

\section{Hydrogen bonding}


The state of hydrogen bonding was investigated by infrared spectroscopy. Spectra provide information about the absorptions from molecular vibrations to determine the conditions of respective bonding. In order to study hydrogen bonds, we focus on the low-energy region from 3000 to 4000 (1/cm). Due to the complex structure of hydrogen bonds in MG, there is very a little literature available to gain information to determine the exact absorption peaks' positions of intra- and intermolecular hydrogen bonding of MG in oil. However due to the physical conditions of MG, the vibration mode of intermolecular hydrogen bonds must be lower than the intramolecular hydrogen bond vibration, so the formation of intermolecular bonds could be distinguished by the shift of infrared absorption peaks.

\begin{figure}
\begin{center}
\resizebox{0.8\textwidth}{!}{\includegraphics{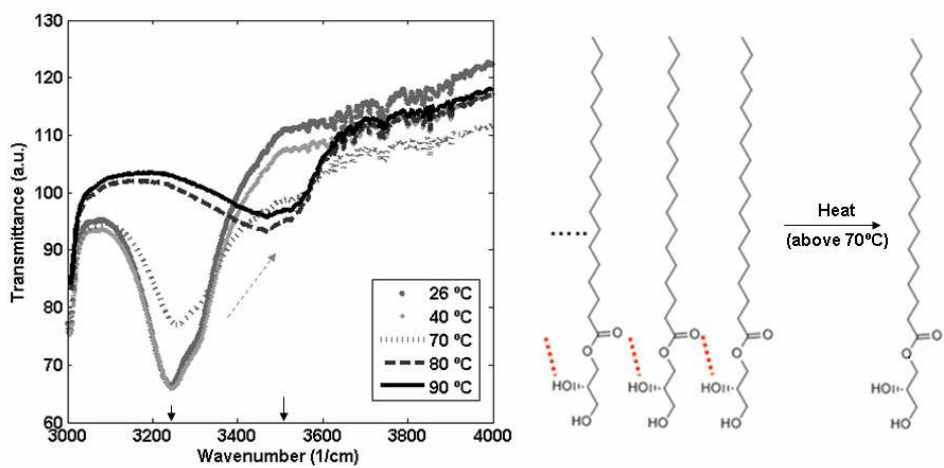}} \caption{The absorption peaks in the infrared spectrum show inter-molecular hydrogen bonds. These bonds are broken by heating and transform into the free hydro-oxide bonding state. The absorption peak therefore shifts from 3250 to 3570 (1/cm)}
\label{fig5}
\end{center}
\end{figure}

Before the measurements the samples were stored at 26$^{\circ}$C over one week to make sure their microstructure was completely aged to the beta crystalline state. Figure \ref{fig5} for the beta-crystal contains an absorption peak at 3250 (1/cm).\cite{book_Dyer69} This peak was suggested to correspond to the intermolecular hydrogen bonding between glycerol groups. When the samples are heated above 70$^{\circ}$C, beta-crystals are melted directly into the isotropic fluid, giving a weak absorption peak at 3570 (1/cm),\cite{book_Dyer69} which characterizes the free hydro-oxide bonding state. This shift clearly shows how the inter-molecular hydrogen bonds is broken by heating above a critical temperature.

\begin{figure}
\begin{center}
\resizebox{0.8\textwidth}{!}{\includegraphics{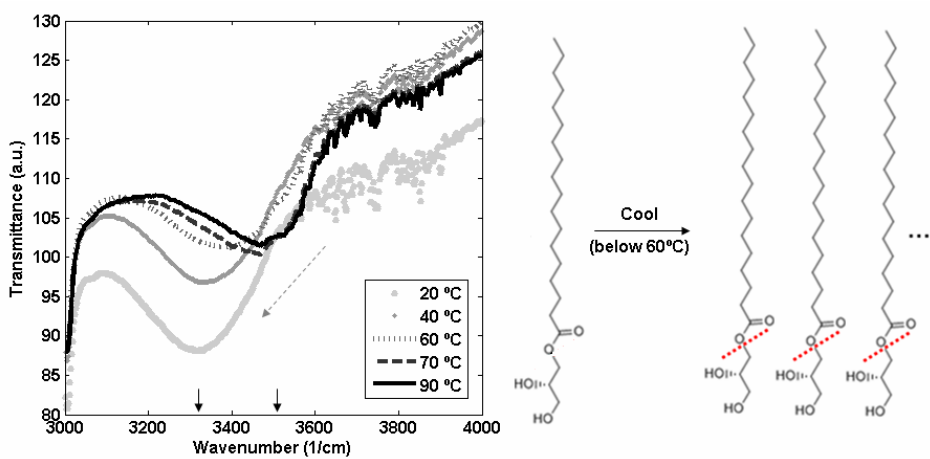}} \caption{On cooling, the hydrogen bonds change their configuration below the gelation temperature (the isotropic-lamellar transition). The intramolecular hydrogen bonds are formed, with the absorption shift from 3570 to 3350 (1/cm). Continuously cooling down from the inverse-lamellar into the sub-alpha crystal phase, we find no change in the state of hydrogen bonding.} \label{fig6}
\end{center}
\end{figure}

The corresponding cooling process was recorded after annealing the solution at high temperature and is shown in Fig.\ref{fig6}. On cooling, the material passed through three phases: isotropic fluid, inverse lamellar and sub-alpha crystal. Below the gelation temperature (50 $^{\circ}$C), surfactants aggregated into the inverse lamellar ordering. Intramolecular hydrogen bonds were formed in a bulk state, characterized by a lower broad band on the infrared spectrum. The absorption peak shifted from 3570 to 3350 (1/cm). By continuously cooling it down to 35 $^{\circ}$C, the material was brought into the sub-alpha crystalline state. However there was no visible change in hydrogen bonding: the infrared spectrum represented the same absorption as in the inverse lamellar phase.

\begin{figure}
\begin{center}
\resizebox{0.9\textwidth}{!}{\includegraphics{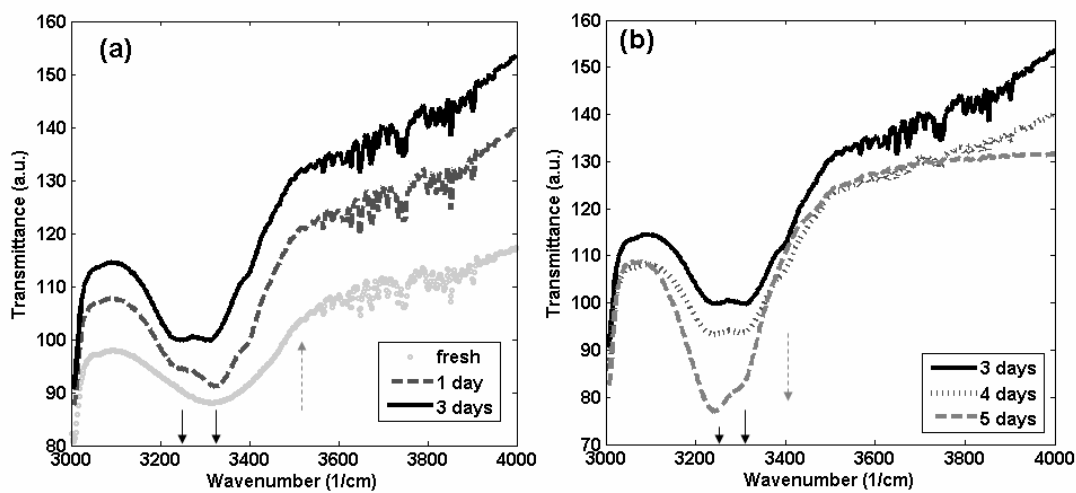}} \caption{(a) During aging, the hydrogen bonding absorption peak gradually splits into two. The upper wavenumber line corresponded to intramolecular bonds, while the lower wavenumber line is related to the lower-energy intermolecular hydrogen bonds.  (b) After five days of aging the ordered intermolecular hydrogen bonds continuously grow and finally begin to dominate.} \label{fig7}
\end{center}
\end{figure}

To gain insight into the connection between aging and the formation of intermolecular hydrogen bonding between MG, the time-evolution of the infrared spectrum was followed at room temperature (26$^{\circ}$C), see Fig.\ref{fig7}. This transformation gradually shows the absorption band splitting into two during aging. The new emerging peak is
attributed to the intermolecular hydrogen bond between alcohol (C-OH) and ester (C=O) groups, respectively. The intermolecular bonding decreases the vibration frequency of molecules so the band shifts to a lower wavenumber at 3250 (1/cm). The coexistence of inter- and intramolecular hydrogen bonding is observed for a long time during aging; the lower-frequency band becomes sharper and increases its intensity, see Fig.\ref{fig7}(b). Over five days, the intermolecular bonding begins to dominate and eventually replaces the intramolecular bonds completely.

We thus conclude that the intramolecular hydrogen bonding (bridging) of glycerol moieties, which establishes when their mobility freezes in the densely packed lamellar state, is not a stable link. The intermolecular hydrogen bonds form and force to rearrange the surfactants in a more ordered way. $D$ and $L$ isomers contain different orientations of hydroxide in the glycerol group. The intermolecular hydrogen bonds between glycerol groups prefer to link the isomers which match the chiral orientation with each other to lower the overall free energy. Therefore $D$ and $L$ isomers spontaneously select the same type of isomers, and this eventually promotes the separation of $D$ and $L$ layers. This reordering process is the key to the aging phenomena we observed on macroscopic scale.

\section{Metastability of inverse lamellar phase}


The observations, and the logic of our arguments about re-configuration of hydrogen bonding being the cause of eventual establishing of the well-ordered beta-crystal form suggests that we examine the aging of the higher-temperature inverse-lamellar phase: from the point of view of infrared absorption there is no difference between it and the fresh sub-alpha phase. We follow the same sequence of experimental techniques.

\begin{figure}
\begin{center}
\resizebox{0.65\textwidth}{!}{\includegraphics{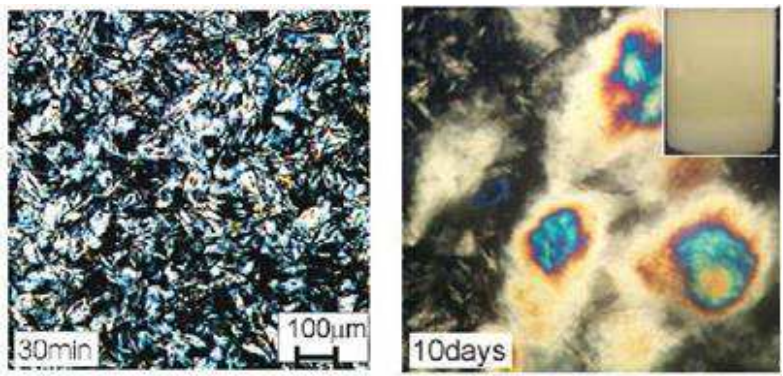}}
\caption{Micrographs of the inverse lamellar phase obtained by polarized microscopy. After 10 days, MG lamellar network would break and form the large structures surrounded by feather-like crystallites. }\label{fig8}
\end{center}
\end{figure}

Figure \ref{fig8} shows the typical microstructure obtained by polarized optical microscopy at fixed temperature (45${}^{\circ}$C), separated by a long period of aging. In order to record clear images, the sample thickness was less than 0.5mm. Similar to the process in the sub-alpha crystalline phase, the samples were annealed at 70${}^{\circ}$C and then cooled down to 45${}^{\circ}$C after which the texture was recorded every 3 hours for two days and subsequently every 12 hours for two weeks. In the inverse lamellar phase MG initially aggregate to form dispersed lamellar bilayer plates which made a percolating network to form a gel and trap the oil inside its structure. This texture shows bright birefringence under the crossed polarizers. The size of every elongated plate domain is roughly $100\mu$m in length and $25\mu$m in width, the same as in the fresh sub-alpha crystalline phase. After 10 days of aging, this network is broken and large agregates (200-600 $\mu$m) with a granular center surrounded by feather-like crystallites were observed.

\begin{figure}
\begin{center}
\resizebox{0.75\textwidth}{!}{\includegraphics{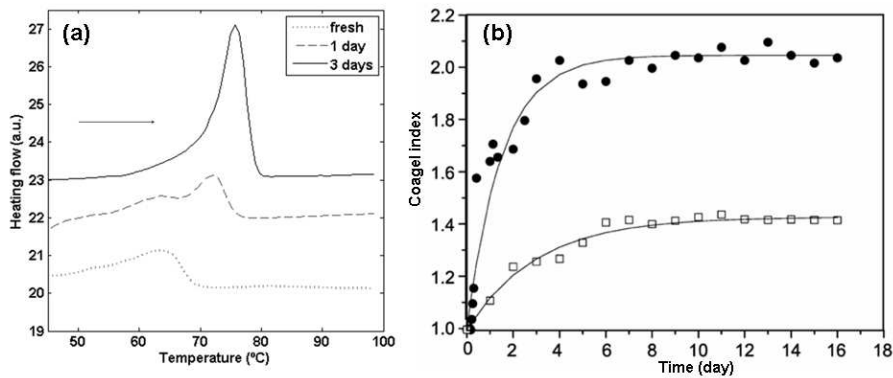}} \caption{(a) During the first two days we find two melting temperatures in the DSC heating scans: the lower one corresponds to the melting of inverse lamellar ordering, while the higher is related to the melting of emerging
beta-crystal. After three days the beta-crystal ordering predominates and the melting enthalpy doubles. (b) In the inverse lamellar phase (45$^{\circ}$C, $\bullet$) CI increased dramatically and then saturated after three days, in contrast to the smaller increase of CI in the aging sub-alpha phase (26$^{\circ}$C, $\Box$). Solid lines represent fitting with the exponential function (see text).} \label{fig9}
\end{center}
\end{figure}

The inverse lamellar ordering with hexagonal head packing of MG/oil mixtures occurred between 37 and 53$^{\circ}$C. The aging was observed in this phase as well and was even faster than the corresponding process in the sub-alpha crystalline state. We attribute the increased aging rate simply to the higher mobility of MG molecules in the lamellar phase. In order to quantify the kinetics of aging, samples stored at 45$^{\circ}$C were used in these experiments. Similar to the procedure used in the sub-alpha crystalline phase, the DSC heating scans were taken from 40 to 100 $^{\circ}$C of the samples with different storage time, see Fig.\ref{fig9}(a). The increase of latent heat was calculated to obtain the evolution of the coagel index, see Fig.\ref{fig9}(b).   This procedure is prone to high errors at the early stages of aging, because we are starting the scan from a temperature that often is in the middle of the second transition, and it is hard to calculate the enthalpy change with precision. Once the two transitions merge after a long period of aging, the calculation of coagel index becomes much more accurate.

The results of infrared spectroscopy revealed that in the early stage of aging, hydrogen bonds drive the restructuring of lamellar bilayers. During aging the sample was partially crystallized and so two melting peaks occurred. The transition at the lower temperature corresponded to the melting of remaining lamellar ordering, while the emerging transition at the upper temperature indicates the melting of the beta-crystals. On aging, the intensity of upper peak increased and the lower peak gradually disappeared. After three days the upper melting transition became dominant. The material was fully aged to the beta-crystalline state, hence the big increase in the coagel index occurred.

To compare the aging process in crystalline and lamellar phases, the evolutions of CI in both phases are plotted in Fig.\ref{fig9}(b). In both cases this increase could be represented by exponential relaxation function: CI$=1+X(1-exp(-t/\tau))$, where $X$ is a fitting constant and $\tau$ the relaxation time. In the sub-alpha crystalline phase, the best fit was achieved with the amplitude $X=0.43$ and the relaxation time $\tau=3$~days. The evolution of the CI in the inverse lamellar phase fit with $X=1.05$ and $\tau=1.5$~days. We conclude that the aging process in the inverse lamellar phase is much faster than in the sub-alpha crystalline state, we assume because the hexagonal heads packing and provided the better chance to rearrange the $D$ and $L$ isomers to reach the beta crystalline state.

Figure \ref{fig9}(b) also shows different shifts of CI during aging in the sub-alpha crystalline and inverse lamellar phases. In the sub-alpha crystalline phase $D$ and $L$ isomers change their arrangement from sub-alpha crystalline to beta crystalline state and the CI increased from 1.0 to 1.4. In contrast, the aging process in the inverse lamellar phase had to experience two steps of ordering. The first step is the crystallization of aliphatic chains in their all-trans conformation, while the second step involves the rearranging of $D$ and $L$ isomers similar to the aging in the crystalline phase. By summing the entropy change from step one and step two, CI increased to 2.0. As expected, during aging of the inverse lamellar phase the increasing of CI is larger than in the sub-alpha crystalline phase.

\begin{figure}
\begin{center}
\resizebox{0.75\textwidth}{!}{\includegraphics{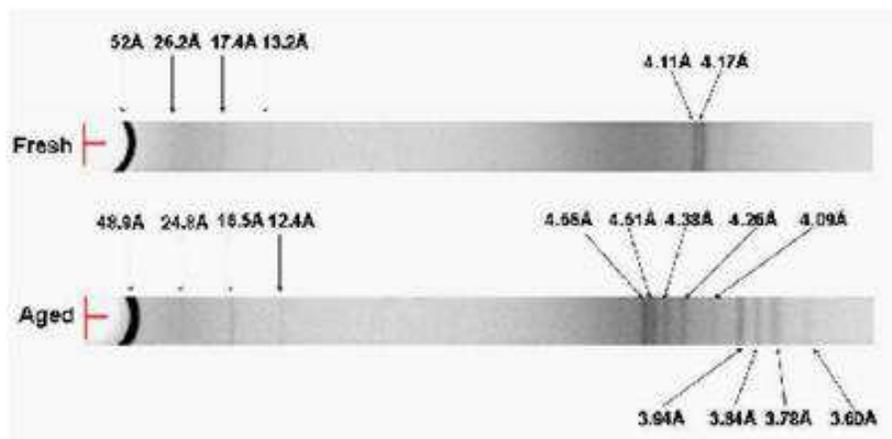}} \caption{Restructuring during the aging of the inverse lamellar phase is monitored by X-ray scattering. In the fresh lamellar phase, the wide-angle peaks show the packing of glycerol groups in a 2-D hexagonal manner.\cite{first} During aging the wide-angle structure breaks into two series of peaks which are identical with the triclinically packed beta-crystal.} \label{fig10}
\end{center}
\end{figure}

In order to describe the aging of micro-structure in the inverse lamellar phase, X-ray diffraction results are shown in Fig.\ref{fig10}. As always, there was a series of concentric rings in the small-angle region to show the existence of lamellar ordering in both fresh and aged materials. The difference between them was revealed by wide-angle diffractions. As expected, the fresh inverse lamellar phase given only twin short spacing X-ray peaks at 4.17\AA and 4.11\AA, characterized the 2-D hexagonal head packing. After three days, the wide-angle scattering pattern developed two series of peaks which was identical with the beta-crystalline state (3-D triclinically arrangement) described in Fig.\ref{fig4} before. This result concluded that the inverse lamellar phase was in fact metastable and eventually transferred to the beta-crystalline state. It means although the aging processes between inverse lamellar and sub-alpha crystalline phases were different, both of them aged to the stable beta-crystalline ordering eventually.

\section{Conclusion \label{sec:6}}

\begin{figure}
\begin{center}
\resizebox{1.1\textwidth}{!}{\includegraphics{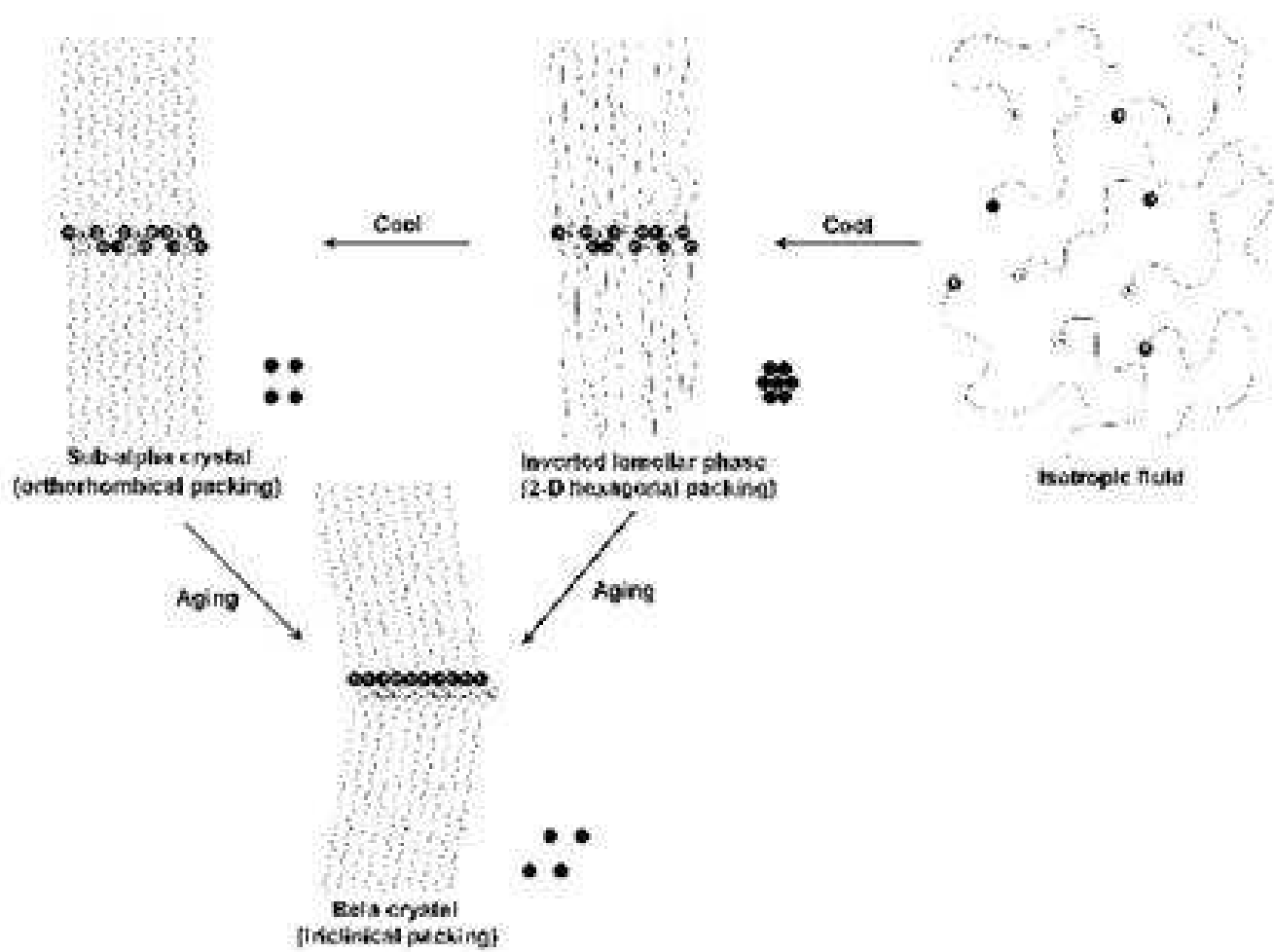}} \caption{Four phases of MG/oil mxtures: isotropic, inverse lamellar (2-D hexagonal head packing), sub-alpha crystalline (3-D orthorhombic packing) and beta crystalline (3-D triclinic unit cell). On aging, intermolecular hydrogen bonds between the polar glycerol head groups are formed and lead to the regular stacking of $D$ and $L$ isomers into the equilibrium structure of beta-crystal. Both inverse lamellar and the sub-alpha crystalline phases are thus metastable and transform into the beta-crystalline state on aging.} \label{fig11}
\end{center}
\end{figure}

The phase behavior of MG/oil mixtures is summarized in Fig.\ref{fig11}. Importantly, all our studies were carried out in the absence of water, which would be expected to have a pronounced effect on intermolecular hydrogen bonding even in small quantities. As reported earlier,\cite{first} there are four generic phases in this system: isotropic, inverse lamellar, sub-alpha crystalline and beta crystalline phases. The study of aging, which takes place in the inverse lamellar and in the sub-alpha crystalline phases, constituted the bulk of this paper. Optical microscopy, calorimetry, X-ray diffraction and infrared spectroscopy results were presented to provide a comprehensive set of both macroscopic and microstructural characteristics of aging.

In order to quantify the aging process, a dimensionless parameter called the coagel index (CI), was defined \cite{Agt98,Veeman05} to represent the relative increase in the order and thermodynamic stability of phases. Here the CI was recorded by DSC to trace the kinetics of aging. In the sub-alpha crystalline phase (at 26$^{\circ}$C) the aging process needed five days to achieve a saturated state of highly-ordered beta crystal in which intermolecular hydrogen bonds were fully established. In order to establish the relationship between aging and the formation of intermolecular hydrogen bonds, the time evolution of infrared spectra had been examined. The coexistence of intra- and intermolecular hydrogen bonding was observed in the initial stages of aging, while the intermolecular hydrogen bonding was dominant and lead to the separation of $D$ and $L$ isomers as a result of aging.

Structural studies confirm the 3-D triclinic packing in the beta-crystalline phase. Two groups of Xray reflections reflect the crystallization of carbon chains and the regular packing glycerol heads, respectively. The separation and reordering of $D$ and $L$ isomers weakenes the emulsifying ability of MG aggregates and causes the metastability of the gel network. The results lead us to the conclusion that both lamellar (2-D hexagonal head packing with a dense fully extended brush of disordered aliphatic chains) and sub-alpha crystalline (3-D orthorhombic packing) phases are in fact metastable, and eventually transformed into the beta-crystalline state (3-D triclinic packing), which has a high melting point. Monoglycerides in oil make a very rare system that forms an elastic gel immediately after ordering from the isotropic phase. Hopefully, the results of this work will help in better understanding of long-term behavior of MG in hydrophobic environment, and improve their use in relevant technologies.

\section*{Acknowledgements}

\thanks{We gratefully acknowledge S.M. Clarke, I. van Damme and A.R. Tajbakhsh for useful discussions and guidance. The help of D.Y.
Chirgadze, in obtaining the SAXS X-ray data is gratefully
appreciated. This work has been supported by Mars U.K.
    }
%

\end{document}